\documentclass[aip,graphicx,sd,reprint,amsmath,amssymb]{revtex4-1}
\usepackage{graphicx}
\usepackage{lineno,hyperref}
\usepackage{epstopdf}
\usepackage{gensymb}
\begin{document}
\preprint{AIP/123-QED}

\title{Suppression of phonon-mediated hot carrier relaxation in type-II InAs/AlAs$_{x}$Sb$_{1-x}$ quantum wells: a practical route to hot carrier solar cells}

%%\author{Elsevier\fnref{myfootnote}}
%%\address{Radarweg 29, Amsterdam}
%%\fntext[myfootnote]{Since 1880.}

\author{H. Esmaielpour, V. R. Whiteside, J. Tang, S. Vijeyaragunathan, T.D. Mishima, S. Cairns, M.B. Santos, and  I.R. Sellers}
\email[I. R. Sellers]{sellers@ou.edu}
\affiliation{Homer L. Dodge Department of Physics \& Astronomy, University of Oklahoma, 440 W. Brooks St., Norman, Oklahoma 73019, USA}

\author{B. Wang}
\affiliation{School of Chemical, Biological and Materials Engineering, Sarkey’s Energy Center, University of Oklahoma, East Boyd Street-T301, Norman, OK 73019, USA}

\begin{abstract}
InAs/AlAs$_{x}$Sb$_{1-x}$ quantum wells are investigated for their potential as hot carrier solar cells. Continuous wave power and temperature dependent photoluminescence indicate a transition in the dominant hot carrier relaxation process from conventional phonon-mediated carrier relaxation below 90 K to a regime where inhibited radiative recombination dominates the hot carrier relaxation at elevated temperatures. At temperatures below 90 K photoluminescence measurements are consistent with type-I quantum wells that exhibit hole localization associated with alloy/interface fluctuations. At elevated temperatures hole delocalization reveals the true type-II band alignment; where it is observed that inhibited radiative recombination due to the spatial separation of the charge carriers dominates hot carrier relaxation. This decoupling of phonon-mediated relaxation results in robust hot carriers at higher temperatures even at lower excitation powers. These results indicate type-II quantum wells offer potential as practical hot carrier systems.
\end{abstract}
\pacs{}
\keywords{Hot carriers, Phonon bottleneck effect, Type-II band alignment}

\maketitle

%\linenumbers

\section{Introduction}
Hot carrier solar cells (HCSCs) have been proposed as devices, which can increase the conversion efficiency of a single junction solar cell above the Shockley-Queisser limit  \cite{1,2,3}. Since thermalization of photogenerated carriers is a major loss mechanism in conventional solar cells, HCSCs have the potential to produce higher efficiency devices using simple single gap semiconductor architectures by eliminating the thermal losses associated with electron-phonon interactions    \cite{4,5}. 

However, before their practical implementation can be realized, HCSCs must circumvent two main challenges  \cite{3,4,6}: 1. Find an absorber material in which hot phonons are longer lived than hot carriers such as to provide the required condition to promote reabsorption of these hot phonons, a phonon bottleneck, which significantly reduces hot carrier relaxation through phonon channels  \cite{7}; 2. Implement energy selective contacts     \cite{8,9,10} in which only a narrow range of energy (within the hot carrier distribution) can be extracted, restricting the energy distributed through carriers cooling, therefore minimizing the entropy heat transfer loss \cite{3,4,5}. 

Here, InAs/AlAs$_{0.16}$Sb$_{0.84}$ quantum-wells are investigated as a candidate hot carrier absorber. The use of quantum wells also offers the potential to facilitate the development of energy-selective contacts and fast carrier extraction via resonant tunneling from quantum wells (QWs) making them an attractive potential system for HCSCs.

\section{EXPERIMENTAL RESULTS AND ANALYSIS}

\begin{figure}
\includegraphics[scale=0.43]{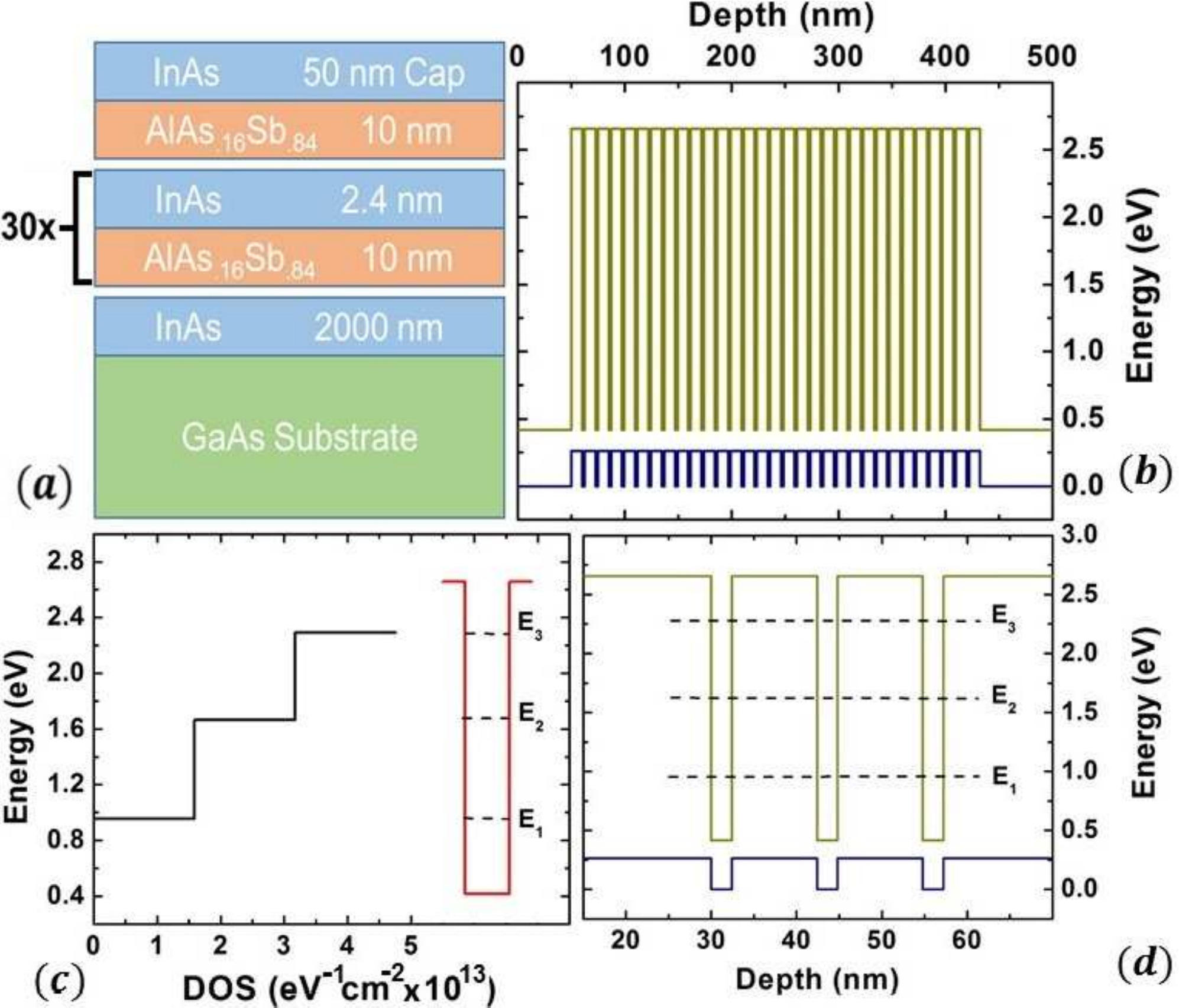}
\caption{\label{figure1}(a) Schematic representation of the InAs/AlAs$_{0.16}$Sb$_{0.84}$ quantum well sample investigated. (b) Simulated energy profiles showing the relative energy of the confinement potential at the Γ point in the conduction (high) and valence band (low). (c) 2D Electron density of states as a function of the energy for this structure. (d) Shows a magnification of the band offsets displaying the type II band alignment with large separation of the energy subbands.}
\end{figure}

A schematic of the sample used in this investigation is shown in Fig. \ref{figure1}(a). The InAs/AlAs$_{0.16}$Sb$_{0.84}$ multi-quantum-well heterostructure was grown by molecular beam epitaxy (MBE) at a substrate temperature of 465\degree{C}. A 2000 nm InAs buffer layer was grown on a nominally undoped GaAs substrate to reduce the density of crystalline defects arising from the lattice mismatch of the active region (MQWs) and the substrate. The thickness of the InAs QWs is 2.4 nm and the AlAs$_{0.16}$Sb$_{0.84}$ barriers are 10 nm. 

As shown in Fig. \ref{figure1}(b), there is a lower quantum confinement in the valence band (VB) and much larger confinement in the conduction band (CB).    The lower energy barrier layers in the VB results in the rapid transfer of the holes absorbed directly in the InAs QWs to the AlAsSb barriers and their enhanced mobility with increasing temperature. Conversely, due to the large confinement in the CB, electrons remain strongly confined at all temperatures    \cite{11,12,13}. The type-II band alignment shown in Fig. \ref{figure1} (b) (magnified in 1(d) for clarity) and thermal diffusion of holes results in an excess of electrons (with respect to holes) in the QWs due to the reduced radiative recombination rate. 

In addition, the large energy band offset between the QW and barrier facilitates absorption of a large proportion of the solar spectrum directly in the InAs QWs, without significant losses in the barriers. Finally, the narrow QWs enable a design in which the separation of the energy levels in the CB is large ($\sim$ 0.7 eV) resulting in a hot carrier distribution that predominately occupies the ground-state subband; without significant influence of broadening due to occupation of higher order (or barrier) subbands (Fig. \ref{figure1}(c), (d)). 

\begin{figure}
\includegraphics[scale=0.83]{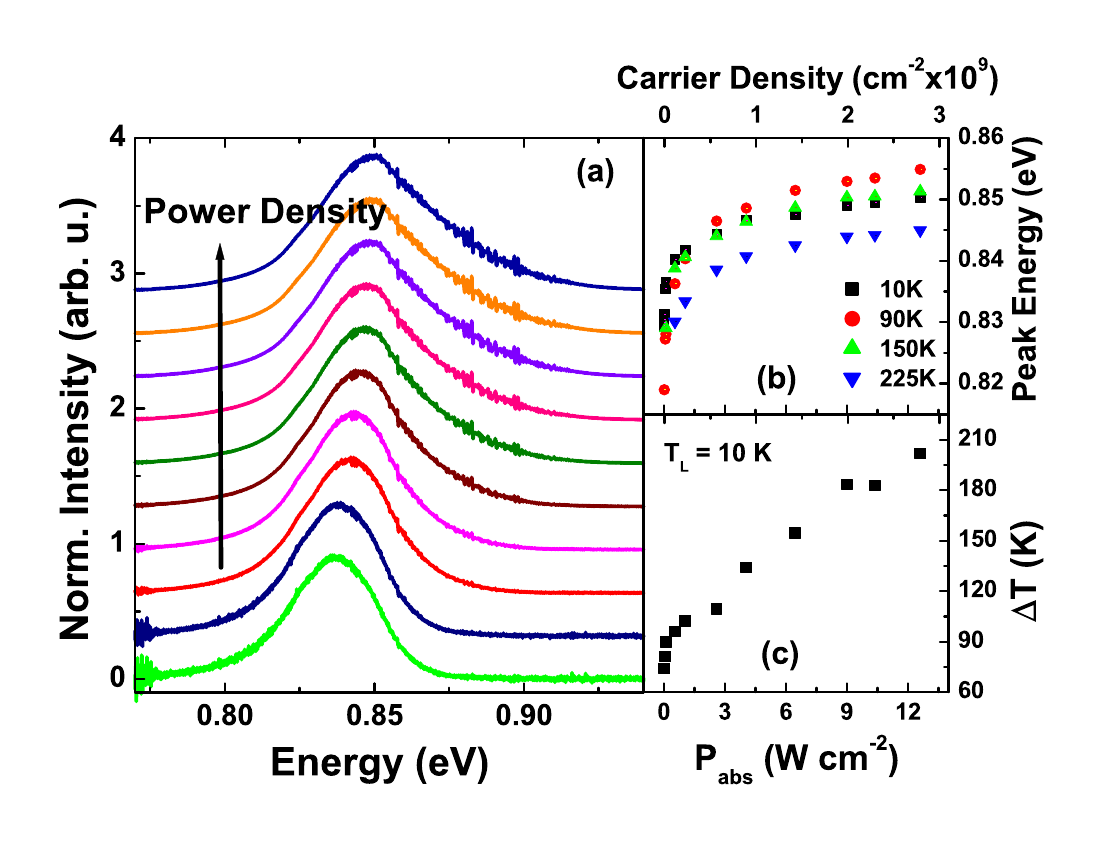}
\caption{\label{figure2}(a) Power dependent PL spectrum at 10 K. (b) Peak PL energy at selected various temperatures. (c) Temperature difference ($\Delta T$) versus absorbed power ($P_{abs}$) at 10 K.}
\end{figure}

Energy dependent photoluminescence (PL) spectra for a range of power densities at 10 K are shown in Fig.\ref{figure2}(a). At lower powers, a shift in the PL peak energy is evident, which reflects the effects of alloy fluctuations that have a significant effect at low power and temperature   \cite{14,15}. At intermediate powers, the peak energy stabilizes and a broadening of the high-energy tail becomes evident. Such high-energy broadening is indicative of the presence of hot carriers \cite{16,17,18} generated by non-equilibrium photogenerated carriers in the CB. 

The observation of the shift in peak energy at low power is also evident in Fig.\ref{figure2}(b), which shows the dependence of the peak PL energy (at increasing temperature) versus absorbed power $(P_{abs})$. At powers below 1-2 W/cm$^{2}$ a large increase in the peak PL is observed. However, at higher $P_{abs}$ the peak PL energy saturates, particularly at higher temperatures. This behavior has been shown to be due to the presence of alloy fluctuations at the InAs-AlAsSb interface and the resulting spatial localization of carriers, which is quenched or saturated, with increasing temperature and/or excited carrier density. \cite{19}

The effect of the alloy fluctuations is also illustrated (somewhat) in Figure 2(c), which shows the power dependent behavior of the temperature difference (the difference between carrier and lattice temperatures: $\Delta T=T_{e}-T_{L}$). The carrier temperature, and therefore, $\Delta T$, can be quantified by fitting the high-energy tail of the PL spectrum, using the generalized Planck relation \cite{12,16,17,18,20,21,22,23}

\small
\begin{equation}
     I(E)\propto\varepsilon(E) exp\left(-\frac{E}{k_{B}T_{e}}\right)
\end{equation}

\normalsize

Where I is the energy-dependent PL intensity, $\varepsilon$ is the effective emissivity, which is related to the absorption profile, $k_{B}$ is Boltzmann's constant, and $T_e$ represents the carrier temperature extracted from the slope of the PL at energy greater than the band gap. Although hot carriers have been predominately investigated using ultrafast time-resolved spectroscopy  \cite{21,22,24}, Equation (1) describes a technique to study the behavior of hot carriers in continuous-wave operation, the mode of operation of solar cells, and therefore presents a more realistic method to interpret the hot carrier dynamics in practical photovoltaic systems. 

At low powers, a large shift of the carrier temperature is observed (below 1 W/cm$^{2}$). The validity of Equation (1) for the extraction of $T_e$ assumes that the effective emissivity $(\varepsilon)$, therefore absorption, is constant at a fixed energy. That is, it is independent of the excitation power. Since the PL energy changes rapidly in the low power regime, the initial increase in $T_e$ is attributed to an artifact of the increasing absorption rather than the real carrier temperature. However (as described above) at higher powers ($P_{abs} >$ 2 W/cm$^{2}$) the energy shift stabilizes (Fig.\ref{figure2}(b)) and as such, reflects the (true) carrier temperature; independent of fluctuation effects, which are saturated under these excitation conditions. 

It is important to emphasize, once more, that since there is a large energy difference between the ground-state transition and the higher-order states (Fig.\ref{figure1}(d)), the high energy tail represents hot carrier effects related solely to the ground-state of the QW, unperturbed by state-filling effects. It must be noted, however, that although band-to-band recombination in the AlAsSb barriers has little effect on the high-energy tail of QW luminescence, the effects of impurities in the QW \cite{25}  and/or localized states at the QW/barrier interface cannot be totally dismissed as contributing to the high-energy tail; the latter of which is discussed in more detail below (see Fig.\ref{figure6}).

The number of carriers generated $(N_c)$ with increasing intensity is indicated with respect to the absorbed power on the upper axis of Fig.\ref{figure2}(b) and (c). The densities absorbed are the order of, or less than, 3$\times$10$^{9}$ cm$^{-2}$ for the excitation levels used. If this is compared to the total 2D density of states calculated for the ground state for the InAs QWs, which is shown in Fig.\ref{figure1}(c) (1$\times$10$^{13}$ cm$^{-2}$), then it becomes clear that the increased contribution of the high energy tail is not the result of significant state-filling, or the saturation of the ground-state, and therefore likely has its origin in inhibited hot carrier relaxation via the creation of a phonon bottleneck \cite{7}. Similar effects have been observed recently in InAs QDs, where the spatial separation of carriers in impurity states leads to the observation of inhibited carrier relaxation as a result of reduced carrier-carrier scattering  \cite{26}. The type-II nature of InAs/AlAsSb is expected to lead to similar results here.

\begin{figure}
\includegraphics[scale=0.25]{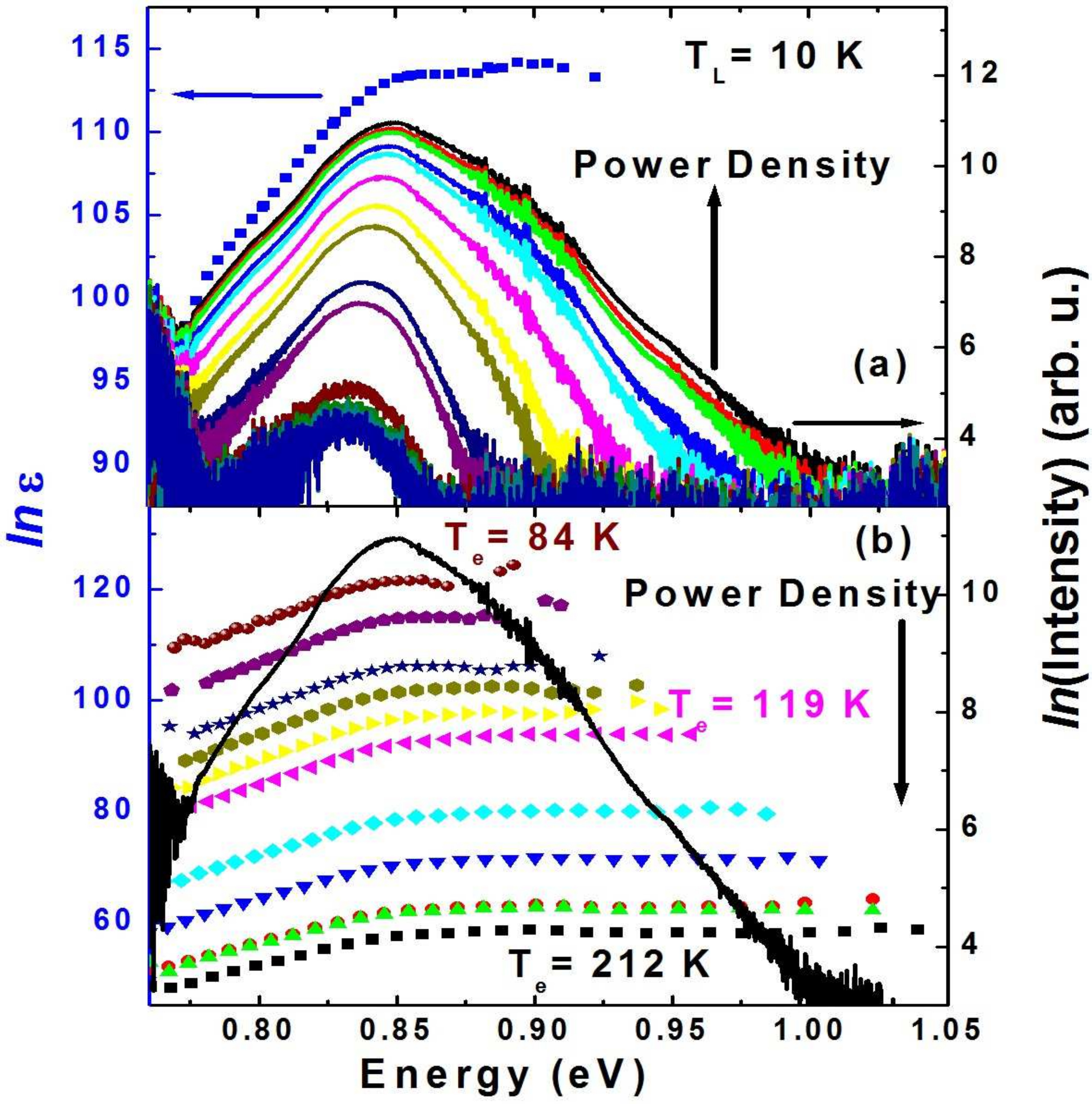}
\caption{\label{figure3}(a) Natural logarithm of effective emissivity (blue squares) and power dependent PL spectrum as a function of energy. (b) The comparison of the behavior of the highest power PL spectrum and natural logarithm of effective emissivity for several intensities.}
\end{figure}

Equation (1) shows that the PL spectrum can also give information about the absorption and the effective band gap of the QWs. The pre-exponential term of the Planck distribution describes an effective emissivity term, which is an energy dependent parameter. Fig.\ref{figure3}(a) shows the natural logarithm of this effective emissivity $(ln \varepsilon)$ (closed squares) as a function of photon energy $(E)$ at 10 K for low excitation power, prior to significant hot carrier generation. These data are shown with respect to the power dependent PL. As the energy increases towards the peak PL energy, and therefore band gap, the effective emissivity increases rapidly. Once the energy gap is reached, the effective emissivity increases much more slowly, reflecting (somewhat) the lower rate of change of the absorption at higher energy \cite{27}.

Fig.\ref{figure3}(b) illustrates the effect of the emissivity term with increasing excitation power obtained by plotting $ln \varepsilon$ $vs.$ $E$ for the PL spectra of Fig.\ref{figure3}(a). These data are compared to highest power PL (shown in black). The behavior of the effective emissivity data is constant across the spectra with the shift in absolute value related to the increasing carrier temperature, as extracted from the slope of the PL, which is inserted in the Eqn. (1) transposed for the natural logarithm of the effective emissivity. The consistency of $ln \varepsilon$ confirms that the pre-exponential term in Equation (1) is indeed independent of power (for the conditions used to extract $T_e$) and demonstrates further that a large separation exists between the ground and the first excited state in the QWs. Therefore, since the carrier temperature is extracted in a region with the constant effective emissivity over a large energy range, $T_e$ is determined with relatively low uncertainty.

\begin{figure}
\includegraphics[scale=0.83]{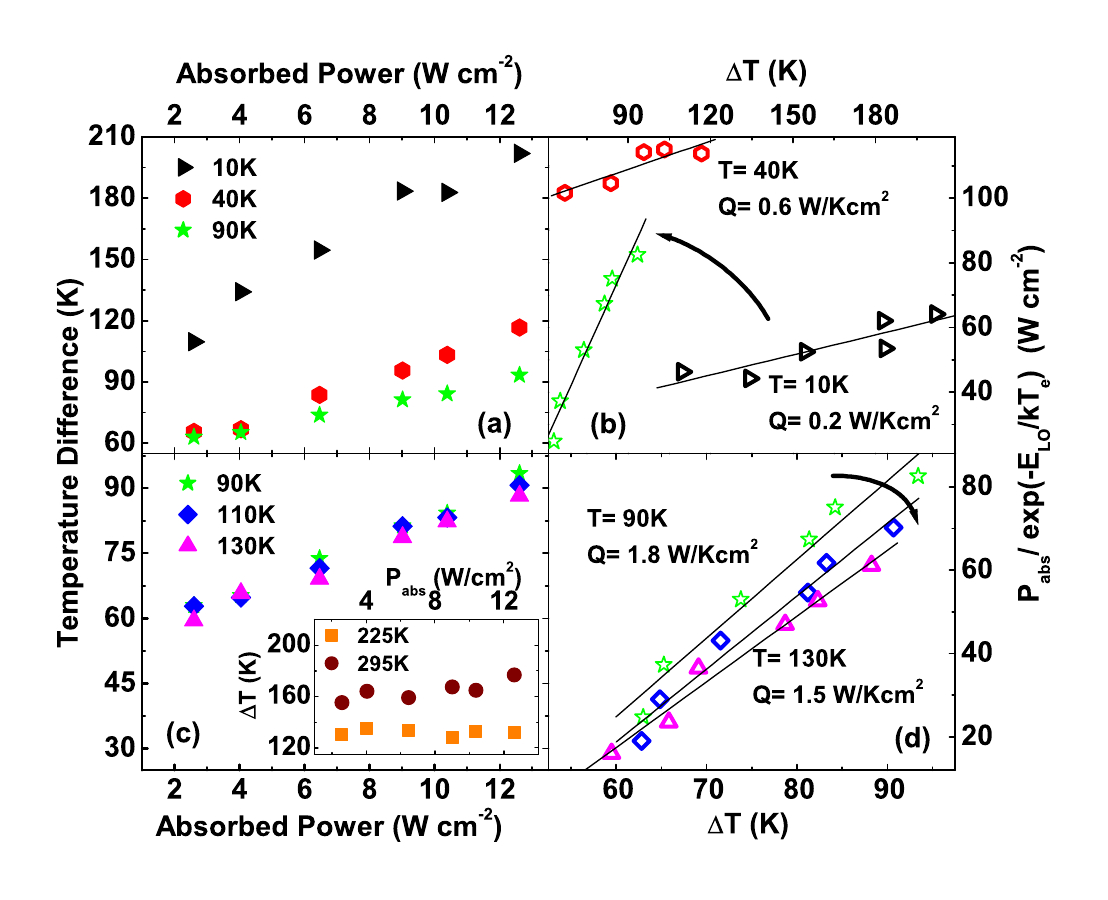}
\caption{\label{figure4}(a), (c)  $\Delta T$ versus power density for several lattice temperatures. (b), (d) Gradients of $P_{abs}$/exp(-E$_{LO}$/(k$_B$T$_{e}$ )) against $\Delta T$ give the thermalization coefficient $(Q)$. The inset graph in (c) displays the independency of $\Delta T$ from power densities at temperatures above 200 K.}
\end{figure}

Fig.\ref{figure4}(a) and (c) display the dependence of the temperature difference $(\Delta T)$ for temperatures between 10 K and 90 K, and 90 K and 130 K, respectively. The inset to Fig.\ref{figure4}(c) shows the same data at 225 K and 295 K. In Fig.\ref{figure4}(a), the carrier temperature, and therefore $\Delta T$, tends to increase with increasing excitation power. The dependence of the hot carriers and their thermalization rate can be evaluated by studying the rate of the thermalized energy (which is the same as absorbed power in the $V_{oc}$  condition) per degree of temperature change \cite{16,17,18} as described by:

\small
\begin{equation}
\noindent P_{th}=\frac{ntE_{LO}}{\tau_{th}}exp\left(-\frac{E_{LO}}{k_{B}T_{e}}\right)=Q\Delta T exp\left(-\frac{E_{LO}}{k_{B}T_{e}}\right)
\end{equation}

\normalsize
Where $P_{th}$ is the thermalized (absorbed) power, $n$ is carrier density, $t$ is thickness, $\tau_{th}$ is the thermalization time, $E_{LO}$ is the phonon energy for InAs, $k_{B}$ is Boltzmann’s constant, and $T_e$ is the carrier temperature. $\Delta T$ is the difference in temperature between the carriers and the lattice, and $Q$ is the thermalization coefficient \cite{16}.

Equation (2) can be used to extract $Q$, an empirical parameter used to assess the contribution of phonon mediated carrier relaxation in QWs \cite{16,17,18}. A high $Q$ is indicative of efficient phonon-mediated relaxation of hot carriers; therefore, systems with lower $Q$ are desired for practical HCSCs \cite{18}. In Fig.\ref{figure4}(b) and (d), the slope of each data set is used to extract $Q$ for that particular temperature. In Fig.\ref{figure4}(b) it can be seen that increasing the temperature to 90 K results in a $Q$ that is increasing, consistent with the increasing contribution of LO phonon scattering at elevated temperature \cite{16}. As the temperature is increased further (90 K - 130 K), as shown in Fig.\ref{figure4}(d), $Q$ starts to become less dependent on $T_{L}$; stabilizing between 90 K and 130 K despite increasing phonon densities at elevated temperature. To reveal the mechanism for this unusual behavior the effect of $T_e$ with increasing excitation power at higher temperatures needs to be considered.

Fig.\ref{figure4}(a) shows $T_e$ versus power between 10 K and 90 K. As excitation power is increased the carrier temperature also increases, as the ratio of excited carriers to phonon density becomes larger. As the lattice temperature is increased to 90 K, the absolute increase in $T_e$ (with power) begins to slow and reduces. This behavior is expected since the phonon density is larger at elevated temperatures (see Fig.\ref{figure6}b), increasing the prevalence of carrier thermalization. This behavior consequently leads to an increased $Q$ as observed in Fig.\ref{figure4}(b). However, for higher temperatures ($>$ 130 K), we can see the dependence of $T_e$ with absorbed power is less than the dependence of $T_e$ at lower temperature (Fig.\ref{figure4}b).  Indeed, although $T_e$ is reduced up to 90 K - stabilizing somewhat through 130 K rather than producing the expected equilibrium carrier distribution via strong LO phonon-relaxation, $T_e$ actually increases; again, despite an increasing phonon density at elevated lattice temperatures. 

In addition to this apparent decoupling of the phonon-relaxation channels above 130 K, the effect of excitation temperature, also, becomes less pronounced at higher temperature. The inset to Fig.\ref{figure4}(b) shows the power dependence at 225 K (solid squares) and 295 K (solid circles), respectively. What is evident is: that the absolute $T_{e}$ increases relative to $T_L$ above 130 K and from 225 K to 295 K, the carriers are ``hot", even at lower excitation levels. 

This behavior presents an interesting question with respect to the validity of using analysis of $Q$ in type-II systems. The empirical parameter $Q$ has been used previously to assess, or qualify the contribution of phonon-relaxation channels in type-I QWs and evaluate their potential for applications as the absorber in HCSCs \cite{18,28}. Indeed recently, this analysis has also been extended to determine the absolute efficiency that may be produced if such systems were applied to HCSCs under concentrated illumination \cite{18}. 

This analysis, however, is based on two principles: 1) that at high temperatures the dominant relaxation channels are related to LO phonon scattering and 2) a constant carrier temperature with respect to excitation power, occurs at (and represents) the equilibrium condition; i.e., the carriers are thermalized at $T_L$.  In the case of the type-II QWs investigated here, the behavior of the system is not consistent with these assumptions, particularly at $T > 130$ K. The high (and increasing) $T_{e}$, at $T > 130$ K, along with the relative insensitivity to excitation power, suggests the relaxation of hot carriers in this system is \textit{not dominated by electron-phonon interaction.}

\begin{figure}
\includegraphics[scale=0.83]{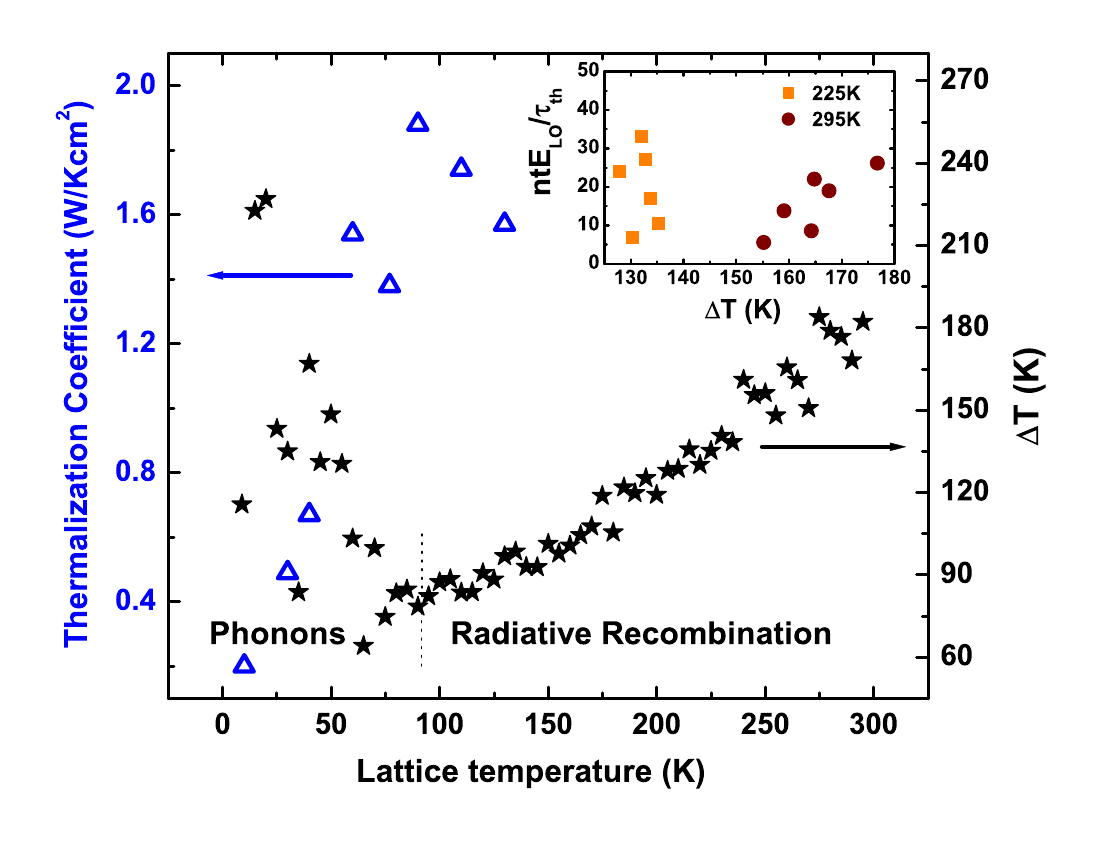}
\caption{\label{figure5} Relation between $Q$ and $\Delta T$ against lattice temperature are displayed to $T_{e}  = 130$ K (blue triangles). The inset shows that $Q$ cannot be determined through for these data sets. Also shown (closed stars) is the temperature dependent hot carrier temperature, $\Delta T$.}
\end{figure}

\begin{figure}
\includegraphics[scale=0.12]{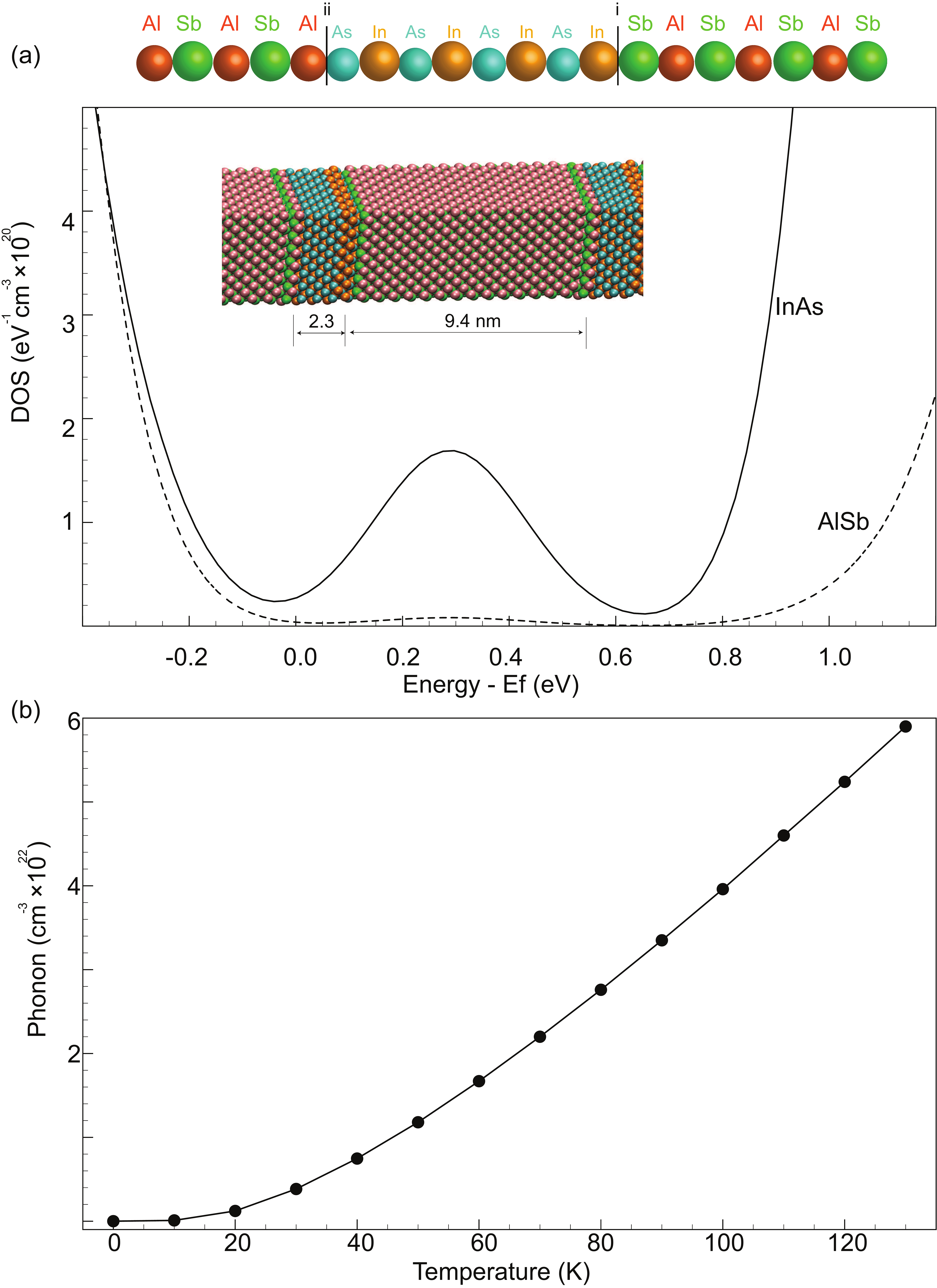}
\caption{\label{figure6} (a) Calculated 3D electron density of states (DOS) for an InAs/AlSb heterostructure, shown inset (upper). The DOS of InAs and AlSb are plotted by projecting the total DOS onto each component. (b) Calculated 3D phonon density of InAs as a function of temperature. The atomic stacking is schematically illustrated in (a) to show the two distinct interfaces. The size of the atoms is shown based on their covalent radius.}
\end{figure}

This is further illustrated in the inset to Fig.\ref{figure5}, which shows the $Q$ analysis at 225 K (closed circles) and 295 K (closed squares). Here, the difficulty in interpreting a thermalization coefficient becomes clear since the independence of $T_e$ with power at these temperatures results in a $Q$ that is large, sometimes infinite, but can also (dependent upon fitting methodology) produce a negative value!

To understand the apparent anomalies in the system under investigation with respect to previous systems presented in the literature \cite{16,17,18}, the nature of the band alignment should be considered. The type-II nature of the InAs/AlAs$_{x}$Sb$_{1-x}$ QWs introduces important differences in the behavior of the samples at high temperature and under intense illumination. At low excitation and at temperatures below 90 K, the PL measured is dominated by a quasi-type-I transition. This is related to recombination of electrons confined in the QWs and holes localized at the InAs/AlAsSb interface \cite{11,12,13}.
At T $>$ 90 K the holes localized at the QW/barrier interface are thermally activated and redistribute into the lower energy AlAsSb barrier region. This delocalization of trapped charges reveals the true type-II band alignment of this system, and consequently the excitons will be spatially separated. It should be noted, if the alloy fluctuations were eliminated, or the materials properties improved, the type-II behavior would be observed at all temperatures. 

A consequence of the separation of the electrons and holes is a reduced radiative recombination efficiency, and therefore a longer radiative lifetime. This behavior will result in an excess of electrons in the QWs, reduced carrier-carrier scattering \cite{24}, and (once more) the development of a phonon bottleneck \cite{7,22,23}. As such, the dominant relaxation process in the type-II QWs presented appears related predominantly to the radiative recombination lifetime, rather than phonon mediated processes.

Relation between $Q$ and $\Delta T$ against lattice temperature are displayed to $T_e$  = 130 K (blue triangles). The inset shows that $Q$ cannot be determined through for these data sets. Also shown (closed stars) is the temperature dependent hot carrier temperature, $\Delta T$.

This behavior further illustrates that the analysis of a thermalization coefficient $(Q)$ used for type-I systems \cite{16} appears invalid here.  Indeed, since the \textit{rapid} spatial separation of carriers absorbed directly in the QWs is a general feature of type-II systems, the decoupling of LO-phonons via inhibited radiative recombination should be a general feature across other type-II QWs investigated this way.    

Fig.\ref{figure5} illustrates further the unique difference between the dominant hot carrier relaxation processes in type-I and type-II systems. Specifically, Fig.\ref{figure5} shows a comparison of the change of $Q$ (open triangles) and $\Delta T$ (closed stars), versus lattice temperature, $T_{L}$. These data are extracted as in a similar manner to those in Fig.\ref{figure4}. At $T < 90$ K, where the sample displays type-I behavior, ΔT decreases with increasing lattice temperature, i.e., the hot carriers are being thermalized by conventional LO-phonon interaction. In this temperature regime (10 K – 90 K) $Q$ is shown to increase with temperature from 0.2 WK$^{-1}$cm$^{-2}$ to 2 WK$^{-1}$cm$^{-2}$, supporting the idea that $Q$-analysis is valid in this regime, when the system behaves as a (quasi)-type-I QW \cite{29}. 

It must be noted, however, that the $Q$ determined here should be considered an upper limit since the diffusion (and therefore mobility) of the carriers absorbed is temperature dependent. Practically, this will result in a change of the absorbed power density as lateral carrier diffusion (and luminescence density area) increases at higher temperatures, before radiative recombination occurs. 

At T $>$ 90 K, the nature of the system changes: as the holes delocalize from alloy fluctuations, the system transitions from (quasi)-type-I to type-II. As such, $T_e$ begins to increase, increasing linearly with increasing lattice temperature, up to 300 K. At temperatures between 130 K and 150 K the behavior, or interpretation, of $Q$ becomes ambiguous as the dependence of the hot carriers with excitation power becomes less pronounced (See Fig.\ref{figure4}(c)). In this regime, $T_e$ is dominated by the efficiency of the radiative recombination, which in type-II systems has been shown to extend for 100’s of nanoseconds \cite{30}. Therefore, the analysis of $Q$ at T $>$ 130 K, or more generally in type-II systems, becomes moot. 
 
    To investigate this hypothesis further, first-principles density-functional-theory (DFT) calculations using the VASP package \cite{31} were used to explore the electronic structure of an analogous InAs/AlSb heterostructure in which the InAs layer is 2.4 nm thick and AlSb is about 9.4 nm thick.  A PBE-GGA exchange-correlation functional  \cite{32} was used for the structural relaxation; when calculating the density of states, a hybrid functional was used  \cite{33,34}. The heterostructure is very similar to that studied experimentally and allows a qualitative picture of its behavior to be determined. In this theoretical system, AlSb, rather than AlAsSb, is used to simplify the interpretation. 

      Fig.\ref{figure6}(a) shows the 3D density of states (DOS) calculated for the structure, which is magnified about the energy gap. The valence band edges of InAs and AlSb in this heterostructure are almost degenerate, while the conduction bands are well separated between InAs and AlSb. The calculations thus support the type-II band alignment. It should be noticed that the band gap is normally underestimated in DFT-PBE calculations and also hybrid-functional calculations. \cite{35}

The first peak above the Fermi level at 0.3 eV is an interfacial state is mainly located at the InAs/AlSb interface, which may account for the carrier localization and the aforementioned transition from type-I to type-II in these QWs. This heterostructure displays two distinct interfaces, i.e. an AlSb-InAs (interface \textit{i}) and an Sb-Al/As-In (interface \textit{ii}) with the difference arising from the varied stacking sequence. A close inspection of the (3D) DOS by projecting it to each atom suggests that the interfacial state is more pronounced at the interface \textit{i}, particularly on the interfacial In, As, and Sb atoms. The origin of these interfacial states is under further investigation but may originate from the interfacial strain effect (Fig.\ref{figure6}(a)). 

      The results indicate that reducing the amount of the Sb (that is, deposit more Al and As) at the interface between the QWs and the barriers may help to reduce these charge trapping levels. The available (3D) DOS of this interfacial state is in the order of 10$^{20}$ cm$^{-3}$ (or 10$^{13}$ cm$^{-2}$ assuming one-nm-thick 2D-interfaces), similar to the InAs conduction band edge. On the other hand, the (3D) phonon density (Fig.\ref{figure6}(b)), that is the overall phonon density without distinguishing different types of phonon, is much higher than the electron density and increases rapidly as a function of temperature. The increased phonon density at higher temperature and the experimentally observed reduced thermalization suggests that phonon-mediated carrier relaxation does not dominate at high temperatures, which is consistent with the hypothesis that a phonon bottleneck results in the type-II QWs presented, supporting the conclusion that the relaxation of hot carriers is dominated by the reduced radiative efficiency in these systems. 

The demonstration of robust hot carriers at elevated temperatures (and at reasonable excitation densities), coupled with the relaxation of phonon loss channels, indicates that type-II systems offer a viable route to practical hot-carrier solar cells.

The InAs/AlAsSb system, specifically, has several attractive features making it a leading candidate: 1) The large QW to barrier energy separation, which is tunable across the solar spectrum, facilitates efficient absorption of the sun’s energy; 2) The degeneracy of the valence band enables efficient hole extraction, while resonant tunneling structures are a reasonable route for fast – energy selective - hot electron extraction in these systems; and 3) Since the photogenerated carriers absorbed directly in the InAs QW are rapidly separated by the type-II band alignment, the loss of photogenerated carriers to photoluminescence is minimized in the QWs. Work is now underway to develop device architectures to further evaluate these systems in practical solar cell devices.

\section{acknowledgments}
The authors would like to acknowledge the contribution of James Dimmock of Sharp Laboratories of Europe Ltd for useful discussions and the critical reading of this manuscript, and Professor Rui Yang (University of Oklahoma) for his insight into the early sample design. The computation was performed at the Extreme Science and Engineering Discovery Environment (XSEDE) and the OU Supercomputing Center for Education \& Research (OSCER) at the University of Oklahoma.

\section*{References}

\bibliography{mybibfile}

\begin{thebibliography}{35}
\bibitem [{29()}]{29}%
  \BibitemOpen
  \bibfield  {title} {\enquote {\bibinfo {title} {Although the behavior of the
  thermalization coefficient observed has qualitatively the behavior expected,
  since the intercept of the linear fit does not pass through zero for all
  temperatures (as expected from eqn. 2) then the quantitative behavior of the
  systems is in fact more subtle, even at low temperatures. this is attributed
  to the quasi-type-i nature of the system, and the contribution of alloy
  fluctuations at lower temperatures.}}\ }\href@noop {} {\ }\BibitemShut
  {NoStop}%
\end{thebibliography}

\end{document}